\documentstyle[12pt]{article}
\newcommand{\be}{\begin{equation}}
\newcommand{\ee}{\end{equation}}
\begin{document}
\begin{center}
{\Large\bf Generalised Hamiltonian embedding of the Proca model}\\[1cm]

N. Banerjee and R. Banerjee\\[.5cm]
S. N. Bose National Centre for Basic Sciences\\
DB 17, Sector I, SaltLake, Calcutta 700064, India.
\end{center}
\vspace{2 cm}
\begin{abstract}

We convert the second class Proca model into a first class theory by using
the generalised prescription of Batalin, Fradkin and Tyutin. We then show
how a basic set of gauge invariant fields in the embedded model can be
identified
with the fundamental fields in the proca model as well as with the observables
in the St\"uckelberg model or in the model involving the interaction of an
abelian 2-form field with the Maxwell field. The connection of these models
with the massive Kalb-Ramond model is also elucidated within a path integral
approach.
\end{abstract}
\newpage
The generalised canonical formalism of Batalin, Fradkin and Tyutin (BFT)
\cite{r1,r2} is a powerful approach to study the connection among different
theories,
in particular it has been effectively used \cite{r3,r4} to convert second class
systems into true (i.e., first class) gauge theories by extending the phase
space
of the original theory. The partition function is then constructed which
reduces
in a definite (unitary) gauge to the partition function of the starting theory.
It is also possible to compute the partition function for other choices of
gauge.
By virtue of the Fradkin-Vilkovisky \cite{r5} theorem, the theories
corresponding
to these partition functions (obtained by different gauge fixings) are all
equivalent.
The basic problem in this approach is to make a judicious choice of gauge that
would yield some  physically interesting theory. Indeed, as is quite well
known, the
choice of a viable gauge is a rather tricky and subtle issue \cite{r6}. Apart
from this there is an arbitrariness in the manner in which the phase space can
be extended. For example, by a straightforward (conventional) extension of the
phase space of the proca model one can obtain, in Faddeev-Popov like gauges the
St\"uckelberg-embedded form of this model \cite{r4}. If, on the contrary, one
adopts an alternative way of enlargement \cite{r7} whereby antisymmetric tensor
fields are introduced, the St\"uckelberg scalar is replaced by 2-form gauge
field. Consequently an equivalence between apparently distinct models may be
achieved since these have a common origin.

The purpose of this paper is to systematically develop the canonical formalism
such that the above mentioned ambiguities related to either gauge fixing or the
embedding procedure are completely avoided. We present our ideas in the context
of Proca model which has received considerable attention \cite{r4,r7}. The
general ideas of BFT will be used to convert the second class constraints and
Hamiltonian of the Proca model into their corresponding first class forms, by
extending the phase space. It is then shown how by performing an inverse
Legendre transformation in this extended space, the St\"uckelberg embedded
Lagrangian of the Proca model emerges naturally. Going back to the extended
phase space, we show that the gauge invariant fields in this space are
equivalent
to the fundamental fields in the Proca model. They satisfy the same algebra and
have identical equations of motion. Furthermore, the involutive Hamiltonian can
be expressed in terms of gauge invariant fields, modulo a term proportional to
the generator of gauge transformation. This signalises the existence of an
underlying gauge theory. One is then led in a natural way to the connection
between the proca model and the gauge invariant sector of a model involving a
massless 2-form gauge field interacting with a Maxwell field. It is important
to
point out that gauge fixing is not used at any stage of the analysis. Neither
do we resort to different embedding procedures to obtain our results. We
conclude
by discussing a purely (configuration space) path integral approach which also
does not involve any gauge fixing.

The Lagrangian for the Proca model is given by,
\be
{\cal L} = -\frac{1}{4} F_{\mu\nu}F^{\mu\nu} + \frac{m^2}{2}A_\mu A^\mu
\label{1}\ee
and leads to the field equations,
\be\partial_\mu F^{\mu\nu} = - m^2 A^\nu.\label{2}\ee
It describes a purely second class system with the constraints,
\begin{eqnarray}
\Omega_0 &=& \Pi_0\approx 0\label{3}\\
\Omega &=& \partial_i \Pi^i + m^2 A_0\approx 0\label{4}
\end{eqnarray}
and the canonical Hamiltonian,
\be
H_c = \int d^4x \left[ \frac{1}{2}\Pi_i\Pi_i + \frac{1}{4}F_{ij}F_{ij} -\frac{
m^2}{2}A_\mu A^\mu - A_0 \partial_i\Pi^i\right]\label{5}\ee
where $\Pi_i = -F_{0i}$ is the momentum conjugate to $A^i$.

We next convert the second class system into first class by adopting the basic
ideas of BFT \cite{r1,r2}. The original phase space is enlarged by introducing
a canonical pair of fields $\theta$ and $\Pi_\theta$. Then a new set of first
class constraints can be defined in this extended space,
\begin{eqnarray}
\Omega_0^\prime &=& \Omega_0 + m^2 \theta\label{6}\\
\Omega^\prime &=& \Omega + \Pi_\theta\label{7}
\end{eqnarray}
which are strongly involutive. It is easy to verify that a Hamiltonian which is
in involution with $\Omega_0^\prime$ and $\Omega^\prime$ is given by,
\be
H^\prime = H_c + \int \left(\frac{\Pi_\theta^2}{2m^2} -\frac{m^2}{2}\partial_i
\theta \partial^i\theta + m^2 \theta \partial_i A^i\right)\label{8}\ee
which satisfies the involutive Poisson algebra,
\begin{eqnarray}
\{ \Omega_0^\prime(x), H^\prime\} &=& \Omega^\prime(x)\label{9}\\
\{ \Omega^\prime (x), H^\prime\} &=& 0\label{10}.
\end{eqnarray}
The first class constraints $\Omega_0^\prime$ and $\Omega'$ are the generators
of gauge transformations,
\begin{eqnarray}
\{ A_0, G[\gamma_0]\} &=& \gamma_0\label{11}\\
\{ A_i, G[\gamma]\} &=& -\partial_i\gamma\label{12}\\
\{\theta, G[\gamma]\} &=& \gamma\label{13}\\
\{\Pi_\theta , G[\gamma_0]\} &=& -m^2\gamma_0\label{14}
\end{eqnarray}
where,
\begin{eqnarray}
G[\gamma_0] &=& \int d^4x \gamma_0(x)\Omega_0^\prime(x)\label{15}\\
G[\gamma] &=& \int d^4x \gamma(x)\Omega^\prime(x).\label{16}
\end{eqnarray}

First we show how to obtain the St\"uckelberg form \cite{r4} by performing an
inverse Legendre tranformation,
\be
{\cal L}^\prime = \Pi_0 \dot{A}_0 + \Pi_i\dot{A}^i + \Pi_\theta\dot{\theta}
- H'\label{17}\ee
The momentum $\Pi_0$ is easily eliminated by using the constraint
$\Omega_0^\prime$.
The other two momenta $\Pi_i$ and $\Pi_\theta$ are eliminated by using
Hamilton's
equations of motion,
\begin{eqnarray}
\dot{A}_i= \{A_i,H'\} &=& -\Pi_i + \partial_i A_0\label{18}\\
\dot{\theta}=\{\theta,H'\} &=& \frac{\Pi_\theta}{m^2}\label{19}
\end{eqnarray}
and we find,
\be
{\cal L}' = -\frac{1}{4}F_{\mu\nu}F^{\mu\nu}+ \frac{1}{2}m^2A_\mu A^\mu
+ \frac{1}
{2}m^2\partial_\mu\theta\partial^\mu\theta - m^2 \theta\partial_\mu A^\mu
\label
{20}\ee
which, upto a boundary term, reproduces the familiar St\"uckelberg structure
\cite{r4} with $\theta$ playing the role of the St\"uckelberg scalar \cite{r5},
\be
{\cal L}'= -\frac{1}{4}F_{\mu\nu}F^{\mu\nu} + \frac{m^2}{2}(A_\mu+\partial_\mu
\theta)(A^\mu+\partial^\mu\theta)\label{21}
\ee

We next show how the Proca model gets related to the abelian 2-form gauge
field.
It follows from (\ref{11}) to (\ref{14}) that the following combinations of
fields,
\begin{eqnarray}
F_0 &=& A_0 + \frac{\Pi_\theta}{m^2}\label{22}\\
F_i &=& A_i + \partial_i\theta\label{23}
\end{eqnarray}
is gauge invariant and satisfy the algebra,
\begin{eqnarray}
\{ F_0(x),F_0(y)\} &=& 0\label{24}\\
\{F_0(x),F_i(y)\} &=& \frac{1}{m^2}\partial_i\delta(\vec{x}-\vec{y})
\label{25}\\
\{ F_i(x), F_j(x)\} &=& 0\label{26}
\end{eqnarray}
The above algebra is precisely identical to the Dirac algebra \cite{r4} of the
Proca fields in which case the constraint $\Omega_0$, $\Omega$
(\ref{3}) are strongly
implemented. Moreover, the $F_\mu$-fields satisfy the equation of motion,
\be
\partial_\mu G^{\mu\nu} = -m^2F^\nu + g^{0\nu}\Omega'\label{27}\ee
obtained from the involutive Hamiltonian (\ref{8}) and $G_{\mu\nu}=\partial_\mu
F_\nu-\partial_{\nu} F_\mu$. Thus, modulo a term proportional to the first
class
constraint $\Omega'$, the equations of motion (\ref{27}) are identical to
(\ref{2}).
We therefore conclude that the $F_\mu$ fields in the embedded (St\"uckelberg)
version play the role of the fundamental fields $A_\mu$ in the Proca model.
The next step is to express the involutive Hamiltonian (\ref{8}) in terms of
the
$F_\mu$ fields,
\be
H'=\int d^3x \left[ \frac{1}{2}\Pi_i^2 + \frac{1}{4}F_{ij}^2 +\frac{m^2}{2}
(F_0^2+F_i^2)\right]-\int d^3x A_0(\partial_i\Pi^i+m^2F_0)\label{28}\ee
Observe that all reference to the original canonical pair ($\theta,\Pi_\theta$)
has been eliminated from the involutive Hamiltonian in favour of the $F_\mu$
fields. Furthermore, (\ref{28}) is written in terms of gauge invariant fields
with $A_0$ playing the role of the Lagrange multiplier associated with the
generator of gauge transformations. It is clear, therefore, that this suggests
an alternative way of exposing the underlying gauge symmetry through the pure
Maxwell term (since this part is exactly reproduced in (\ref{28})) plus
something
which interacts with it. The structure of this `remainder'can be recognised
by realising that $F_\mu$ being gauge invariant and divergenceless (which
follows
from (\ref{27}) and (\ref{10})) allows the introduction of an abelian 2-form
gauge field $B_{\mu\nu}$ by,
\be
F_\mu = \frac{1}{2}\epsilon_{\mu\nu\alpha\beta}\partial^\nu B^{\alpha\beta}
= \frac{1}{6}\epsilon_{\mu\nu\alpha\beta}G^{\nu\alpha\beta}\label{29}\ee
\be G_{\nu\alpha\beta} = \partial_{[\nu}B_{\alpha\beta]}\label{30}\ee
This is similar in spirit to the analysis in 2+1 dimensions where the Hopf
term could thereby be introduced in the nonlinear sigma model \cite{r9} or
the equivalence between the Maxwell-Chern-Simons theory and a self dual model
established \cite{r10}. The involutive Hamiltonian (\ref{28}) may now be
expressed as,
\begin{eqnarray}
H' &=& \int\left[\frac{1}{2}\Pi_i^2 + \frac{1}{4}F_{ij}^2 -\frac{m^2}{12}
G_{ijk}G^{ijk}+\frac{m^2}{4}G_{0jk}G^{0jk}\right]\nonumber\\
&-& \int A_0 \left[ \partial_i\Pi^i + \frac{m^2}{2}\epsilon_{ijk}\partial^i
B^{jk}\right]\label{31}\end{eqnarray}
One immediately recognises that Gauss operator associated with the Lagrange
multiplier $A_0$ as that which occurs in the theory of an abelian 2-form field
interacting with the Maxwell field and whose dynamics is governed by the
Lagrangian density \cite{r7,r11},
\be
\tilde{{\cal L}} = -\frac{1}{4}F_{\mu\nu}^2 -\frac{m^2}{6}\epsilon_{\mu\nu
\rho\sigma}
A^\mu G^{\nu\rho\sigma} + \frac{m^2}{12}G_{\mu\nu\rho}G^{\mu\nu\rho} \label{32}
\ee
It takes only a slight effort to show that the complete involutive Hamiltonian
(\ref{31}) follows from (\ref{32}). The canonical momenta obtainable from
(\ref{32})
are given by,
\begin{eqnarray}
\Pi_0=0; & \Pi_{0i} = 0\label{33}\\
\Pi_i = -F_{0i}; & \Pi_{ij} = m^2(\epsilon_{ijk}A^k + G_{0ij})\label{34}
\end{eqnarray}
and the canonical Hamiltonian is
\begin{eqnarray}
\tilde{H} &=& \int d^3x \left[ \Pi_i\dot{A}^i + \frac{1}{2}\Pi_{ij}\dot{B}^{ij}
- \tilde{{\cal L}}\right]\nonumber\\[.5cm]
&=& \int d^3x \left[ \frac{1}{2}\Pi_i^2 +\frac{m^2}{2}A_i^2 + \frac{1}{4}
F_{ij}^2
+\frac{1}{4m^2}\Pi_{ij}^2 + \frac{1}{2} \epsilon_{ijk}A_i\Pi_{jk}\right.
\nonumber\\
&+& \left.\frac{m^2}{12}G_{ijk}^2 - A_0(\partial_i\Pi^i +\frac{m^2}{2}
\epsilon_{ijk}
\partial^iB^{jk}) + B_{0j}\partial_i\Pi_{ij}\right]\label{35}
\end{eqnarray}
Apart from the primary constraints (\ref{33}) there are two secondary
constraints,
\begin{eqnarray}
\partial_i\Pi^i + \frac{m^2}{2}\epsilon_{ijk}\partial^iB^{jk} &\approx & 0
\label{36}
\\ \partial_i\Pi_{ij} &\approx & 0\label{37}
\end{eqnarray}
As expected, all these constraints are first class. Using the definition for
the
canonical momenta $\Pi_{ij}$ (\ref{34}), it is easy to show that (\ref{31})
maps
on to (\ref{35}) modulo a term proportional to the first class constraint
$\partial_i\Pi_{ij}=0$. Since the physical states are annihilated by such
constraints
the equivalence between (\ref{31}) and (\ref{35}) in the gauge invariant sector
is established. We therefore find out that the Proca model is alternatively
mapped on to the gauge invariant sector of models defined by either the
St\"uckelberg Lagrangian (\ref{21}) or the  Maxwell-Kalb-Ramond (MKR)
Lagrangian
(\ref{35}), both of which are derived within a unified canonical framework.

Furthermore
the correspondence among the basic fields in these models is given by,
\begin{eqnarray}
(A_0)_{\rm Proca}\leftrightarrow(A_0+
\frac{\Pi_\theta}{m^2})_{\hbox{St\"uckelberg
form}}\leftrightarrow(\frac{1}{6}\epsilon_{ijk}G^{ijk})_{\rm MKR}\label{38}\\
(A_i)_{\rm Proca}\leftrightarrow(A_i+ \partial_i\theta)_{\hbox{ St\"uckelberg
form}}\leftrightarrow(-\frac{1}{2}\epsilon_{ijk}G^{0jk})_{\rm MKR}\label{39}
\end{eqnarray}
and are the analouges of the corresponding mappings obtained by one of us \cite
{r10} in 2+1 dimensional case.

The Proca model or its  equivalent formulations describe the
propagation of a single massive mode. It is known  \cite{r11}
that this is also true for a model described by a free massive
abelian  2-form potential - the Kalb-Ramond (KR) model whose
Lagrangian is given by,
\be
{\cal L}_{KR} = \frac{1}{12}G_{\mu\nu\rho}G^{\mu\nu\rho} -
\frac{1}{4}B_{\mu\nu} B^{\mu\nu}\label{40}\ee

We shall conclude this paper by explicitly revealing the
connection of (\ref{40}) with the Proca model (\ref{1}) and
thereby with the other alternative formulations (\ref{21}) and
(\ref{35}). To this effect consider the following master
Lagrangian,
\be
{\cal L}_{M} = \frac{m^2}{2}A_\mu A^\mu -\frac{m^2}{4}
B_{\mu\nu} B^{\mu\nu} + \frac{m}{6}
\epsilon_{\alpha\mu\nu\rho}A^\alpha G^{\mu\nu\rho}\label{41}\ee
Master Lagrangians, incidentally, have been exploited earlier
\cite{r12} to discuss the equivalence among various theories but
these were usually confined to 2+1 dimensions \cite{r13}. The
partition function corresponding to (\ref{41}), in the presence
of external sources $J_\mu$, $K_\mu$, is given by,
\be
Z_M = \int {\cal D}A_\mu {\cal D} B_{\mu\nu} \exp i\int d^4x \left(
{\cal L}_M + J_\mu A^\mu + \frac{K_\mu}{6}
\epsilon_{\mu\alpha\beta\sigma} G^{\alpha\beta\sigma}\right) \label{42}\ee
Performing the gaussian integration over the abelian 2-form
field $B_{\mu\nu}$ yields,
\be
Z_M = \int {\cal D}A_\mu \exp i\int d^4x \left(\frac{m^2}{2}
A_\mu A^\mu -\frac{1}{4}F_{\mu\nu}^2 + J_\mu A^\mu +
\frac{1}{m^2} K_\beta\partial_\alpha F^{\alpha\beta}\right)
\label{43}\ee
where a nonpropagating contact term  has been dropped. In the
absence of sources (\ref{43}) is exactly the partition function for
the proca model \cite{r1}.

Alternatively, doing the $A_\mu$-integration in (\ref{42}) leads
to,
\begin{eqnarray}
Z_M &=& \int {\cal D}B_{\mu\nu} \exp i \int d^4x \left( \frac{1}{12}
G_{\mu\nu\rho}G^{\mu\nu\rho} - \frac{m^2}{4}
B_{\mu\nu}B^{\mu\nu} \right.\nonumber\\[.1cm]
&+& \left.\frac{1}{6}\epsilon_{\mu\alpha\beta\sigma}G^{\alpha\beta\sigma}
(J^\mu+K^\mu)\right)\label{44}\end{eqnarray}
where, once again, a nonpropagating contact term has been
ignored. In the absence of sources (\ref{44}) is the partition
function for the Kalb-Ramond model (\ref{40}). Since (\ref{43})
and  (\ref{44}) were derived from a common origin, it
establishes their duality. Furthermore, comparing the source
terms proportional to $J_\mu$, $K_\mu$, the following
identifications are obtained,
\be
(A_\mu)_{\rm Proca}\leftrightarrow
(-\frac{1}{m^2}\partial^\alpha F_{\alpha\mu})_{\rm Proca}
\leftrightarrow
(-\frac{1}{6m}\epsilon_{\mu\alpha\beta\sigma}G^{\alpha
\beta\sigma})_{ \rm KR}\label{45}\ee
It is straightforward to reproduce the equations of motion of
either the Proca field or the Kalb-Ramond field from the above correspondence.
This correspondence can also be used, in conjunction with
(\ref{38}), (\ref{39}) to relate the basic fields in the
St\"uckelberg and MKR versions with those in the KR model.

To conclude, we have employed the general notions of Batalin,
Fradkin and Tyutin \cite{r1,r2} to obtain a deeper insight into
the connection between the fields of the Proca model and those
in the corresponding St\"uckelberg embedded version or in a
model involving the interaction of a massless abelian 2-form
field with the Maxwell field (Maxwell-Kalb-Ramond model). A
significant aspect of this work was to provide a unique
enlargement technique of the phase space of the original Proca
model  which led, in  a systematic and natural way, to the gauge
invariant sector of either the St\"uckelberg type model or the
MKR model. The use of different enlargement prescriptions, as
advocated in the literature \cite{r1,r2,r7}, can therefore be avoided.
Moreover, since gauge fixing was not necessary at any stage of
the computations, subtleties and ambiguities involved in such a
procedure were eliminated. We also furnished a path integral
formulation, whereby starting from a master Lagrangian the
duality between the Proca model and the massive Kalb Ramond
model was established. An identification between the fields in
the respective models was also obtained. The master Lagrangian
was gauge fixed from the begining so that the problem of gauge
fixing was once again bypassed. Although the computations were
presented in 3+1 dimensions, it is easy to extend these to
arbitrary $d+1$ ($d\geq 3$) dimensions. The Proca field in that
case would be connected to a $d-1$-form abelian field. The $d=2$
example, incidentally, is special since it admits the Chern
Simons 3-form and has been discussed extensively in the
literature \cite{r12,r14}.

\end{document}